\documentclass[12pt]{article}
\usepackage{epsfig,graphicx}

\title{Magnetic moments of mesons}
\author{ A.M.Badalian, Yu.A.Simonov\\
State Research
Center\\Institute of Theoretical and Experimental Physics, \\
Moscow, 117218 Russia}

\newcommand{\beq}{\begin{eqnarray}}
 \newcommand{\eeq}{\end{eqnarray}}
\newcommand{\be}{\begin{equation}}
 \newcommand{\ee}{\end{equation}}

\def\ga{\mathrel{\mathpalette\fun >}}
\def\fun#1#2{\lower3.6pt\vbox{\baselineskip0pt\lineskip.9pt
\ialign{$\mathsurround=0pt#1\hfil ##\hfil$\crcr#2\crcr\sim\crcr}}}

\newcommand{{\SD}}{\rm SD}

\newcommand{\vesig}{\mbox{\boldmath${\rm \sigma}$}}

\newcommand{\veP}{\mbox{\boldmath${\rm P}$}}
\newcommand{\vep}{\mbox{\boldmath${\rm p}$}}

\newcommand{\vez}{\mbox{\boldmath${\rm z}$}}
\newcommand{\veS}{\mbox{\boldmath${\rm S}$}}
\newcommand{\veL}{\mbox{\boldmath${\rm L}$}}

\newcommand{\vel}{\mbox{\boldmath${\rm l}$}}
\newcommand{\veR}{\mbox{\boldmath${\rm R}$}}

\newcommand{\veta}{\mbox{\boldmath${\rm \eta}$}}
\newcommand{\veB}{\mbox{\boldmath${\rm B}$}}

\newcommand{\veJ}{\mbox{\boldmath${\rm J}$}}

\newcommand{\vepi}{\mbox{\boldmath${\rm \pi}$}}

\newcommand{\vemu}{\mbox{\boldmath${\rm \mu}$}}

\newcommand{\lan}{\langle}
\newcommand{\ran}{\rangle}

\begin{document}
\maketitle
\begin{abstract}
Magnetic moments of charged and neutral mesons are calculated with the use of
the relativistic Hamiltonian derived from the path integral form of the
$q_1\bar q_2$ Green's function. The magnetic moments are shown to be expressed
via the average quark energies which are defined  by the fundamental
quantities: the string tension $\sigma$, the current quark masses, and  the
strong coupling constant $\alpha_s$. Resulting values for vector, axial, and
tensor light and $K$ mesons agree well with all available lattice data.

\end{abstract}

\section{Introduction}

Magnetic moments of hadrons are important dynamical
characteristics, which can be useful in many areas, e.g for
nucleons, and can give new information on the hadron dynamics,
being also a serious test of the dynamics put in a model. In this
paper we shall exploit the QCD dynamics in the form of the
relativistic Hamiltonian (RH) of a meson , derived from the QCD
path integral \cite{ref.1}, which was already extensively used in
the studies of hadron spectra without external fields
\cite{ref.2}. The extension of the RH to the case of external
magnetic field (MF) has been done recently in Ref.~\cite{ref.3},
where the meson spectrum as a function of MF was calculated.

Within this method the magnetic moments of baryons have also been
studied analytically in Ref.~\cite{ref.4} and for the lowest
hyperons their values, calculated in the first approximation,
agree with experiment within 10\% accuracy. (The same accuracy was
achieved for the baryon magnetic moments before, in
Ref.~\cite{ref.5}, where the QCD string dynamics was exploited
with approximated wave function).

At the same time for last decades a thorough analysis of the
hadron magnetic moments is also being done in the lattice QCD (see
\cite{ref.6} for the review).

In this paper we are mainly interested in the magnetic moments of vector,
axial, and tensor mesons and to this end, develop further the method of
Ref.~\cite{ref.4}, suggested for baryons. For the sake of generality, here we
shall use the RH in MF, both for charged and neutral mesons, from
Refs.~\cite{ref.3,ref.7} and in this way define all terms in the expansion of
the hadron mass in powers of MF. (This is done in Section 2). We show that the
meson magnetic moments are easily calculated through the average quark energies
$\omega_i$, which here are defined by the minimal set of the QCD fundamental
parameters: the string tension $\sigma$, $\alpha_s$, and the current quark
masses, not introducing any fitting parameters.

Having done our analytical calculations, in Section 3 we compare
our results for light and $K$ mesons with the lattice calculations
of the meson magnetic moments \cite{ref.8}, \cite{ref.9} and find
a good agreement within the accuracy of the lattice and our
calculations. Discussion of the results and prospectives are given
in Section 4.

\section{The Hamiltonian for a meson in magnetic field}

The path-integral Hamiltonian for the $q_1\bar q_2$ system in MF
was derived in Ref.~\cite{ref.3}, \cite{ref.7} and has the form,
\be H=H_0 + H_\sigma +W,\label{1}\ee where \be H_0 =\sum^2_{i=1}
\frac{\left(\vep^{(i)} - \frac{e_i}{2} (\veB\times
\vez^{(i)})\right)^2+ m^2_i +\omega^2_i}{2\omega_i},\label{2}\ee
\be H_\sigma =- \frac{e_1\vesig_1\veB}{2\omega_1}
-\frac{e_2\vesig_2\veB}{2\omega_2},\label{3}\ee

In (\ref{1}) the term $W$ contains the confinement potential
$V_{conf}$, the  perturbative gluon-exchange (GE) potential
$V_{GE}$, and spin-dependent interaction $V_{SD},$ as well as the
nonperturbative self-energy term $\Delta M_{SE}$ \cite{ref.10},

\be W=V{\rm conf}+V_{OGE}+V_{SD}+\Delta M_{SE}. \label{4}\ee All
these terms have been introduced and extensively studied in case
without MF in Refs.~\cite{ref.2}, \cite{ref.10}-\cite{ref.12}.

In (\ref{2}),(\ref{3}) the following basic elements of the
path-integral approach enter: the average energies $\omega_i$,
which play actually the role of the einbein parameters
\cite{ref.13}, being  defined from the eigenvalues
$M_n(\omega_1,\omega_2)$ of the Hamiltonian  $H$. \be
H\Psi=M_n\Psi,\label{5}\ee using the stationary point equations:
\be \frac {\partial
M_n(\omega_1\omega_2)}{\partial\omega_1}|_{\omega_i=\omega_i^{(0)}}=0,~~
i=1,2.\label{6}\ee As a result
$M_n(\omega_1^{(0)},\omega_i^{(0)})$ is  our prediction for the
mass of a given meson.

In the case without MF the eigenvalues $M_n$ were already
calculated for all kinds of mesons: light-light \cite{ref.2} ,
heavy-light \cite{ref.14}, heavy quarkonia \cite{ref.15}, and in
all cases good agreement with experiment was obtained. Note, that
$M(\omega_i^{(0)}\omega_i^{(0)})$ are the functions of the current
quark masses, the string tension $\sigma$, and the strong coupling
$\alpha_s(r)$, i.e. do not contain any fitting parameters, in
contrast to other relativistic model approaches.

Then we introduce the c.m. and relative coordinates of the $q_1
\bar q_2$ system, \be \veR =
\frac{\omega_1\vez^{(1)}+\omega_2\vez^{(2)}}{\omega_1+\omega_2},~~\veta=
\vez^{(1)}-\vez^{(2)},\label{7}\ee and also make an ansatz for the
wave function, \be \Psi(\veta,\veR)=\exp(i\Gamma) \varphi (\veta,
\veR),\label{8}\ee where $\Gamma=\veP\veR-\frac{\bar
e}{2}(\veB\times \veta)\veR$. Then one can get the Hamiltonian
$H'_0$, acting on $\varphi(\veta, \veR)$. If the c.m. motion is
chosen to associate with the total charge of the meson, equal
$e_1+e_2.$, then one has to put $\bar e=\frac{e_1-e_2}{2}$ and the
new Hamiltonian $H'_0$ is obtained in the form,
$$ H'_0=\frac{\veP^2}{2(\omega_1+\omega_2)}+\frac{(\omega_1+\omega_2)\Omega^2_R\veR^2_\bot}{2}
+\frac{\vepi^2}{2\tilde \omega}+ \frac{\tilde \omega
\Omega^2_\eta\veta^2_\bot}{2}+X_{LP}\veB\veL_P+$$
\be+X_{L_{\eta}}\veB\veL_\eta+X_1\veP(\veB\times\veta)+X_2(\veB\times
\veR)(\veB\times\veta)+ X_3\vepi(\veB\times
\veR)+\frac{m^2_1+\omega^2_1}{2\omega_1}+
\frac{m^2_2+\omega^2_2}{2\omega_2}.\label{9}\ee

Here $\veL_\eta=(\veta\times\frac{\partial}{i\partial\veta}),~~
\veL_P=(\veR\times\frac{\partial}{i\partial\veR}), ~~ \tilde
\omega=\frac{\omega_1\omega_2}{\omega_1+\omega_2}$. All coefficients $X_i$ are
given in Appendix 1, while $\Omega_R,\Omega_\eta$
 are following, \be\Omega^2_R= B^2\frac{(e_1+e_2)^2}{16\omega_1\omega_2}\label{10}\ee
\be \Omega^2_\eta=\frac{B^2}{2\tilde \omega(\omega_1+\omega_2)^2}
\left[ \frac{(e_1\omega_2+\bar
e\omega_1)^2}{2\omega_1}+\frac{(e_2\omega_1-\bar e\omega_2)^2}
{2\omega_2}\right].\label{11}\ee

Our purpose here is to study the first order corrections $O(eB)$,
proportional to spin and the angular momentum $\veL_\eta$,  to the
meson mass $M(\omega_1^{(0)}, \omega_2^{(0)})$ in the total
expansion of the meson mass  $M(B)$ in powers of the MF $B$ (here
the relation $\veL_P\equiv 0$ is assumed):

\be M(B)= M(0) - \vemu_S\veB + X_{L_{\eta}}\veL_{\eta}\veB +
\sum^{\infty}_{n=1}\kappa_n B^n . \label{12}\ee

Taking into account (\ref{9}), (\ref{10}), (\ref{11}),  one
arrives at the mass formula for a meson in the form, $$
M(B)=\frac{P^2_z}{2(\omega_1+\omega_2)}+ \Omega_R (2n_{R\bot}+1)+
$$

\be \left\lan\frac{\tilde \omega\Omega^2_{\eta}\veta^2_{\eta\bot}}{2}\right\ran
+ \frac{m^2_1+\omega^2_2}{2\omega_1} + \frac{m^2_2+\omega^2_2}{2\omega_2} -
\frac{e_1\vesig_1\veB}{2\omega_1}-\frac{e_2\vesig_2\veB}{2\omega_2}
+X_{L\eta}\veL_{\eta}\veB +\lan\Delta M_X\ran.\label{13}\ee In (\ref{13}) the
term $\Delta M_{X_i}$ implies the sum of all terms with coefficients
$X_i~~(i=1,2,3)$.

Now, expanding (\ref{13}) in powers of $B$, one can write
($\veL_{\eta}\equiv \vel$) \be  M(B) =  M(0) -\vemu\veB+ B\kappa_1
+B^2\kappa_2+X_l\vel\veB+O(B^3),\label{14}\ee where \be \vemu =
\frac{e_1\vesig_1}{2\omega^{(0)}_1(B=0)}+\frac{e_2\vesig_2}{2\omega_2^{(0)}
(B=0)}-X_l\vel.\label{15}\ee Here the term $B\kappa_1$ can be
obtained, expanding $\omega_i^{(0)}(B)$ in $B$ and keeping the
first order term, but all these terms do not contribute to the
magnetic moments. Also in (\ref{13}) $\lan\Delta M_X\ran$
contributes only to terms $O(B^2)$, if the vector state considered
has no  internal angular momenta $\veL_P$.  In this case all terms
in $\Delta M_X$ vanish in the first order, $(\Psi\Delta
M_X\Psi)=0$. As a result magnetic moments of mesons  in (\ref{15})
are calculated through only $B=0$ characteristics of the $q\bar q$
system, namely only $\omega_i^{(0)}(B=0)$ are needed, while
magnetic polarizabilities require higher terms in $(eB)$ from
(\ref{13}).

\section{Results for  vector, axial, and tensor mesons}

The expression (\ref{15}) allows to calculate the magnetic moments of mesons
with different quantum numbers, both for the angular momentum $l=0$ and $l\neq
0$. Below we perform calculations for the light and $K$ mesons, while in
similar way the magnetic moments of  heavy-light and heavy-heavy  mesons can be
also defined.

a) The case of zero internal angular momentum, $l=0$\\

The absolute value $\mu_S$ of the magnetic moment $\vemu =\mu_S\veS$ for the
$S$-wave mesons with $l=0$ can be obtained, taking $S_z=\frac12(\sigma_{1z}+
\sigma_{2z})=+1$, so that one can write
\be\mu_S=\frac{e_1}{2\omega_1^{(0)}(B=0)}+\frac{e_2}{2\omega_2^{(0)}(B=0)},\label{16}\ee
where $e_1(e_2)$ refers to the charge of quark $q_1$ (antiquark $\bar q_2$).
Both $\omega_i^{(0)}(B=0)$ are supposed to be found from the same Hamiltonian
(\ref{1}) with $B=0$, using as the only input $\sigma, \alpha_s(r)$, and the
current quark masses $m_i$ ; their values are discussed in details in Appendix
2.

The values of $\omega_i^{(0)}(B=0)$ for $\rho$ and $K^*$ were taken from
\cite{ref.2}, \cite{ref.16} and listed in Tables I, II. Notice that the values
of $\omega_i^{(0)}$ given  there  refer  to the case when GE interaction with
the standard form of $\alpha_s(r)(n_f=3)$ is taken into account, while in the
absence of GE potential their values for the ground states of light and $K$
mesons would be by $\sim 15\%$ smaller (see (A2.14)).

The most simple case refers to $\rho^\pm$, when neglecting quark
masses  $(m_u=m_d=0),$ the following values were calculated:
$\omega_1^{(0)}(B=0)=\omega_2^{(0)}(B=0)\equiv \omega_0 =0.397 $
GeV. Then in nuclear magnetons (n.m.) \be \mu_S(\rho)\equiv
\mu_{\rho} =\frac{e}{2\omega_0}=\frac{M_P}{\omega_0}~({\rm
n.m.}).\label{17}\ee

In this case the magnetic moment, $\mu_\rho(1S)=2.37$ n.m.,
appears to be large  (see Table I) and close to that calculated on
the lattice in Refs.~\cite{ref.8}, \cite{ref.9}, where
$\mu_\rho(1S, lat)\simeq 2.4$ n.m. In the same way  for higher
radial excitations the magnetic moments $\mu_\rho(2S)$ and
$\mu_\rho(3S)$, given in last column of Table I, are obtained.
Note, that in our calculations we have neglected the S-D mixing of
excited $\rho$ states, as well as the influence of the spin-spin
interaction $V_{ss}$ on the value of $\omega_0$, i.e. $V_{ss}$ is
considered as the first order correction to the $\rho$ mass, which
is defined via $\omega_0$.

The same procedure is used for the $K^{*\pm}$ mesons, for which
important values are given in Table II. For $K^*$ a close
agreement with lattice data from Refs.~\cite{ref.8}, \cite{ref.9}
also takes place  (see Table III).

Of special interest are  the magnetic moment of neutral vector
meson $K^{*0}$. As it follows from (\ref{15}), if $ e_1=-e_2$,
then the magnetic moment is proportional to the difference
$(\omega_1^{(0)}-\omega_2^{(0)})$, which in its turn is
proportional to $(m^2_1-m^2_2)$ and vanishes, when the current
masses of the strange quark and d-quark are taken to be equal,
$\tilde m_s=m_d$. For $K^{*0} (d\bar s)$, using (\ref{16}) with
$e_1=-\frac{e}{3}, e_2=+\frac{e}{3}$ and taking $\omega_1
=\omega_n, \omega_2 =\omega_s$   for $m_d=0, m_s =0.2$~GeV from
Table II, one obtains the magnetic moment of $K^{*0}(d\bar s):$

  \be \mu (K^{*0})= - 0.0972 e({\rm ~GeV})^{-1}=-0.183~ {\rm n.m.},\label{18}\ee
which is much smaller than the magnetic moment of $K^{*+}$. The
same result is clearly seen in lattice  data \cite{ref.8}, where
for a neutral meson the linear dependence of its magnetic moment
on the squared mass $m^2_\pi$, which is proportional to $m_q$, is
observed, thus corresponding to small magnetic moment.
\vspace{1cm}

b) {\bf The case of nonzero $l$} \vspace{0.5cm}

Firstly one can consider  the simple  situation, when $\veS=0,~
\vel \neq 0$ and in this case clearly $\veJ =\vel$ and $\vemu =
\mu_l \veJ.$  Hence, one has a simple correspondence:
 \be \bar \mu =\mu_l \cdot \frac{2M_p}{e} ~({\rm n.m.}), ~ ~\mu_l\equiv
 -X_{l\eta} = \frac{ e_1\omega^2_2 + e_2\omega^2_1}{2\omega_1\omega_2
(\omega_1+\omega_2)}.\label{19}\ee  The values of $\bar \mu$ for
$l=1,2$ are given in Tables II, III.

In  another case, if $\veS =1$, one should define the average
value of $\vemu$ in the situation, when $\veJ$ is the only vector
of the system, as

\be \bar \mu = \lan J, J_z| \mu_z | J, J_z\ran ; ~~J_z =J;~~ \bar
\mu=\sum_{m_s,m_l} | C^{JJ}_{Sm_S, lm_l}|^2 (\mu_S m_S+\mu_l m_l),\label{20}\ee
where $\vemu$ is given by \be \vemu = \mu_S \veS + \mu_l \vel, ~~ \mu_S =
\frac{e_1}{2\omega_1} + \frac{e_2}{2\omega_2};\label{21}\ee and $\veJ = \vel
+\veS$. As a result,  one obtains the averaged magnetic moment in units
$e$(GeV$^{-1})$,
  \be \bar \mu = \frac{\mu_S+\mu_l}{2} (J=l=S=1), ~~ \bar \mu =
  \frac{3\mu_l-\mu_S}{2}~~
  (J=1, l=2, S=1),\label{22}\ee where $\mu_S, \mu_l$ are given in (\ref{21}) and (\ref{19}),
respectively.

Then from Eq.~(\ref{22}) one can easily find magnetic moments of all meson
states with $l=1,2$ and $S=1$; they are given in Table 1 for light mesons and
in Table 2 for strange mesons.

These results refer to the positive (negative) charge mesons,
while for neutral light mesons with $\omega_1 = \omega_2$  their
magnetic moments are identically zero. For the strange neutral
mesons, as mentioned above, their magnetic moments are
proportional to $(\omega_1 -\omega_2)\sim m^2_1 - m^2_2$ and
therefore they are also very small. While using corresponding
values of $\mu_S, \mu_l$ with $e_1 =-e_2$ from (\ref{19}),
(\ref{21}), one can find the values of $\bar \mu$ for any neutral
meson with $S=1$, using (\ref{22}). It is important that the
Eq.(\ref{20}) is applicable for the states with all possible
values of $S, l,$ and $J$.

\section{Discussion  of results and conclusions}

The main result  of our study  is the expression (\ref{20})  for the magnetic
moment of an arbitrary meson, expressed through the factors $\mu_S, \mu_l$ and
hence, through the averaged quark energies $\omega_1, \omega_2$. The latter are
calculated here within the same relativistic Hamiltonian (see Appendix 2),
which contains only the first principle input of QCD: current quark masses,
string tension, and $\alpha_s$ in coordinate space. In this way magnetic
moments of mesons in the $LS$ scheme, neglecting the $L, L\pm 2$ mixing, were
calculated. This mixing can be easily included within our method, provided the
mixing amplitudes are known from experiment ($e^+e^-$ cross sections of
mesons), or from theoretical models.

Our main results for the light and $K$ mesons are presented in Tables 1,2 and
in Tables 3-5 they are compared  to other calculations. The most significant
our result is for the mesons with the spin $S=1$ (see Table 3), where the
magnetic moments of different mesons are compared to all existing lattice
calculations, since both approaches are of the first principle QCD
calculations.

From  Table 3 one can see an encouraging agreement within the accuracy of the
lattice calculations for all mesons considered. This agreement can be further
detailed, using dependence of the averaged energies $\omega_i$ on varying
current quark masses: it can be translated in the lattice study of the magnetic
moment dependence on the pion mass squared -- see a quantitative analytic
analysis of this dependence in a recent publication \cite{ref.17}.

In Tables 4,5 calculated magnetic moments are compared to the existing model
calculations: the sum rule approach \cite{ref.18}-\cite{ref.20}, the
constituent quark model \cite{ref.21}, the Dyson-Schwinger approach
\cite{ref.22}. One can see a reasonable agreement only for the $S$-wave mesons,
like $\rho^+, K^{*+}$; however, in other cases a strong disagreement is
obtained and this implies that the sum rule method is less reliable for higher
excitations of mesons. Another line of possible development is the calculation
of magnetic moments for growing MF, in  which case for some mesons, the
energies $\omega_i$ are growing like $\sqrt{ eB}$ and therefore their magnetic
moment decrease. For other mesons, having different spin projection on the MF,
their $\omega_i$ decrease  \cite{ref.3}, \cite{ref.7} and in this case their
magnetic moments are growing with the MF.

 As it is, our results give an additional support for the
relativistic Hamiltonian approach,  generalized here to the case of arbitrary
strong magnetic and electric fields as in Ref.~\cite{ref.3}. We have shown that
the varying quark energies $\omega_i$ and their final stationary point values
$\omega_i^{(0)}$, used in our approach, yield the relevant physical
information, which can be applied in different directions: to calculate meson
masses with and without magnetic field,  the meson magnetic or electric moments
of different mesons in strong magnetic field. Our results on magnetic moments
can be easily generalized  to the case of heavy-light or heavy-heavy mesons.

One should stress that our approach, which is quite successful
also in case of magnetic moments, considers only valence part of
the wave functional,  while sea quark part is ignores. One might
expect, that the additional $q\bar q$ components are important for
high excited mesons near decay thresholds, and   they can be
accounted for by the multichannel formalism.

  The same method
can be used to study other relativistic systems of two or more
constituents: atoms or positronium, different hadrons, and also
nuclei in strong MF. The relevant physical situation may exist in
astrophysics (magnetars) and in colliding ions, as discussed in
\cite{ref.3}.

 The authors are grateful for useful discussions to M.A.Andreichikov and
 B.O.Kerbikov.


\begin{table}
\caption{The light meson magnetic moments $\bar \mu ~~(\sigma
=0.18 $ GeV$^2$, $m_u=m_d=0$, $\omega(nl)=\omega_1=\omega_2)^{a)}$
\label{tab.1}}

\begin{tabular}{|l|l|l|}
\hline\hline
    meson state  & $\omega(nl)$ (GeV)&$\bar \mu$ (n. m.)\\\hline

    1S & 0.397&2.37\\\hline
    2S&0.549&1.71\\\hline

    3S&0.667&1.41\\\hline
    1P &0.489&$a_1^+(^3P_1)~~1.44;~~ a^+_2(^3P_2)~~ 2.88$\\
&&$b(^1P_1)~~ 0.96$\\\hline

2P &0.616&$a_1^+~~1.14; ~~a^+_2~~ 2.28$\\
&&$b_1(^1P_1) ~~ 0.76$\\\hline

1D &0.571& $a^+_3 ~~3.29;~~ \rho(^3D_1)~~ 0.411$\\\hline

2D &0.681& $a^+_3~~ 2.76;~~ \rho(^3D_1)~~ 0.345$\\\hline
     \hline

\end{tabular}

$^{a)}$ The  parameters of $\alpha_{GE}(r)$ are given in Appendix
2, Eqs. (A.2.8)-(A.2.12).
\end{table}

\begin{table}
\caption{ K meson magnetic moments ($m_n=0, m_s=0.20~$GeV$, \sigma=0.18$
GeV$^2$, $\alpha_{GE}^{a)}$\label{tab.2}}

\begin{tabular}{|l|l|l|l|}
\hline\hline

state &$\omega_n$~(GeV)&$\omega_s$~(GeV)& $\bar \mu$~(n.m.)\\\hline

    1S & 0.411&0.467&219\\\hline
    2S&0.559&0612&1.73\\\hline

    3S&0.676&0722&1.36\\\hline
    1P &0.500 &0.544 &$K^{*0}(^3P_2),~-0.10$\\
&&&$K^{*}_1(^3P_1),~1.38$\\
&&&$K^{*}_1(^1P_1),~0.93$\\
&&&$K^{*}_2(^3P_2),~2.76$\\\hline

2P &0.625&0.667&\\
\hline

1D &0.580 & 0.617&$K^{*}(^3D_1),~0.405$\\\hline

2D &0.689 &0.725&$K^{*}(^3D_1),~0.183$\\\hline
     \hline

\end{tabular}

$^{a)}$ See the footnote to Table 1.
\end{table}

\begin{table}
\caption{Magnetic moments (in n.m.) of the lowest vector and axial mesons in
comparison with lattice calculations \label{tab.3}}

\begin{tabular}{|l|l|l|l|l|l|l|}
\hline\hline
    meson
    &$\rho^+(1S)$&$K^{*+}$&$a^+_1(1^3P_1)$&$K^{*+}_1(^3P_1)$&$\rho_1(^3D_1)$&$K^{*+}(^3D_1)$\\\hline

m.m. (this paper)&2.37&2.194&1.44&1.38&0.411&0.405\\\hline

m.m.$^{*)}$ lattice [8]&2.4&2.4&1.5&1.5&0.5&0.5\\\hline

m.m.$^{*)}$ lattice [9]&2.3&2.1&-&-&-&-\\

\hline\hline
\end{tabular}
$^{*)}$ The values of lattice  magnetic moments are taken from the Figs.
\cite{ref.8},\cite{ref.9} and their accuracy is $\ga 10\%$.

\end{table}

\begin{table}
\caption{Magnetic moments of vector and axial mesons (in n.m.) in
comparison with other model calculations\label{tab.4}}

\begin{tabular}{|l|l|l|l|l|}
\hline\hline

$\rho^+$& $K^{*+}$&$K^{*(0)}$&$a^+_1$& source
\\\hline

2.37&2.19& -0.183& 1.44& this paper \\\hline

$2.4\pm 0.4$& $2.0\pm 0.4$&$-0.28\pm0.04$&$3.8\pm0.6$ &[18] QCD
sum rules\\\hline

1.92& -&- & - &[20] light-front \\\hline
 2.14& -& -&- &[21]  light-front\\\hline 2.2&2.08&-0.08&-&[22] Schwinger-Dyson eq.\\\hline \hline
\end{tabular}

\end{table}

\begin{table}
\caption{Magnetic moments (in n.m.) of the lowest tensor
mesons\label{tab.5}}

\begin{tabular}{|l|l|l|l|l|l|l|}
\hline\hline
    meson
    &$f^t_2$& $f_2^0$& $a^\pm_2$& $a^0_2$&  $K^{*+}_2 (1430)$& $K^{*0}_2(1430)$\\\hline

m.m.  &2.88&0&2.88&0&2.76&-0.10\\ (this paper)&&&&&&\\\hline

m.m.  [19] &2.1$\pm 0.5$& 0& 1.88$\pm$ 0.4& 0& $0.75\pm 0.08$&
$0.076\pm 0.008$\\\hline \hline
\end{tabular}

\end{table}

\newpage
\vspace{2cm}
 \setcounter{equation}{0}
\renewcommand{\theequation}{A1.\arabic{equation}}

\hfill {  Appendix  1}

\centerline{\large \bf Coefficients $X_i, i=L_\eta,LP,1,2,3$}

 \vspace{1cm}

\be
X_{L_\eta}=-\frac{e_1\omega_2^2+e_2\omega^2_1}{2\omega_1\omega_2(\omega_1+\omega_2)}\label{A1}\ee
\be X_{LP} =-\frac{e_1+e_2}{2(\omega_1+\omega_2)}\label{A2}\ee \be
X_1 = \frac{e_2\omega_1-e_1\omega_2-\bar e
(\omega_1+\omega_2)}{2(\omega_1+\omega_2)^2}\label{A3}\ee \be X_2=
\frac{-\omega_2(\bar e-e_1)(e_1\omega_2+\omega_1\bar e)- (\bar
e+e_2) (e_2\omega_1- \bar
e\omega_2)\omega_1}{4\omega_1\omega_2(\omega_1+\omega_2)}\label{A4}\ee
\be X_3= \frac{\bar e(\omega_1+\omega_2)
-e_1\omega_2+e_2\omega_1}{2\omega_1\omega_2}, ~~ \bar e=
\frac{e_1-e_2}{2}.\label{A5}\ee

\vspace{2cm}
 \setcounter{equation}{0}
\renewcommand{\theequation}{A2.\arabic{equation}}

\hfill {  Appendix  2}

\centerline{\large  \bf The Hamiltonian without magnetic fields}

 \vspace{1cm}

If there are no MF $B$, the relativistic Hamiltonian $H_0$ \cite{ref.1}  can be
presented as

\begin{equation}
\tilde H_0=
\frac{\omega_1}{2}+\frac{m_1^2}{2\omega_1}+\frac{\omega_2}{2}+
 \frac{m_2^2}{2\omega_2}+\frac{\vep^2}{2\tilde \omega}+ V_{\rm static} (r) \equiv T_R + V_{\rm stat} (r) ,
\label{eq.01}
\end{equation}
where by derivation the quark mass cannot be chosen arbitrarily and must be
equal to the current mass $\bar m_q$ for the $u,d$, and $s$ quarks. In our
calculations $\bar m_q=0$ for the $u,d$ quarks, $m_s\simeq \bar m_s(1~{\rm
GeV})=200$~MeV for the $s$ quark \cite{ref.23}. Here the mass $m_s$ is larger
than the conventional $\bar m_s(2~{\rm GeV})=95\pm 20$~MeV, taken at the scale
$\mu=2$~GeV, and the reason for that difference originates from the fact that
in the static interaction  the $s$-quark current mass $ m_s(\mu)$ enters at a
smaller scale, $\mu\sim 1$~GeV. It is important that the current quark masses,
used in our relativistic Hamiltonian, allow to avoid such fitting parameters as
the constituent quark masses, usually present in other models.

In the static interaction, $V_{\rm stat}=V_{\rm conf}+ V_{\rm
GE}$, the linear confining potential $V_{\rm conf} =\sigma \cdot
r$ is taken here with the string tension $\sigma=0.18$~GeV$^2$,
which cannot be considered as a fitting parameter, since its value
follows from the slope of the Regge trajectories for light mesons
\cite{ref.2}.

The choice of GE potential is important for low-lying light and
$K$ mesons, while its influence is much smaller for high
excitations. Here we use the vector strong coupling, denoted as
$\alpha_{\rm B}(r)$, which possesses the asymptotic freedom
property and freezes at large distances at the value $\alpha_{\rm
crit}$, and its parameters are not arbitrary, as we show below.

The variables $\omega_i$, entering  RH $\tilde H_0$, have to be
determined from the extremum conditions: $\frac{\partial
H_0}{\partial \omega_i}=0~(i=1,2)$, that gives
\begin{equation}
\omega_i(nl)=\langle\sqrt{ \vep^2+m_i^2}\rangle_{nl} \quad
(i=1,2). \label{eq.02}
\end{equation}
These average  energies, $\omega_1(nl)$ and $\omega_2(nl)$, refer
to the quark $q_1$ and the antiquark $\bar q_2$, while $\tilde
\omega$ is the reduced mass,  $\tilde \omega
=\frac{\omega_1\omega_2}{\omega_1+\omega_2},$ and
$\vel=\vel_1+\vel_2$. Then putting $\omega_i$ into
Eq.~(\ref{eq.01}), one arrives at a different form of the kinetic
energy term, denoted as $T_{\rm R}$:
\begin{equation}
 T_R=\sqrt{\vep^2+m_q^2} + \sqrt{\vep^2+m_c^2}.
\label{eq.03}
\end{equation}
Rigorously, the expression (\ref{eq.03}) for $T_{\rm R}$ is valid
only for $l=0$, while in general case, for $l\neq 0$, $T=T_{\rm
R}+T_{\rm str}$ contains additional kinetic energy term, $T_{\rm
str}$, which appears because, besides a standard rotation of a
quark and an antiquark, the string rotates itself. It was shown in
Refs. \cite{ref.2}, \cite{ref.24} that for $l\leq 4$ the
contribution from $T_{\rm str}$ is relatively small compared to
the e.v. $M_0(nl)$ and therefore $T_{\rm str}$ can be considered
as a perturbation. Still its matrix element (m.e.) $\Delta_{\rm
str}(nL)=\langle T_{\rm str}\rangle_{nl}$ has to be included in
the mass formula of a meson.

Then the e.v. $M_0(nl)$ and the meson w.f. are defined by the
spinless Salpeter equation (SSE):
\begin{equation}
 \left[T_{\rm R} + V_{\rm B}(r)\right]\varphi_{nl}=M_0(nl)\varphi_{nl} .
\label{eq.04}
\end{equation}
However, the spin-averaged meson mass $M(nl)\equiv M_{\rm
cog}(nl)$ includes not only the e.v. $M_0(nl)$ (\ref{eq.04}), but
also two additional negative contributions: the string correction
$\Delta_{\rm str}(nl)=\langle H_{\rm str}\rangle_{nl}$
\cite{ref.2}, if $l\neq 0$, and the nonperturbative self-energy
(SE) term $\Delta_{\rm SE}(nl)$ \cite{ref.10}:
\begin{equation}
 M(nl)=M_0(nl) + \Delta_{\rm str}(nl)+ \Delta_{\rm SE}(nl).
\label{eq.05}
\end{equation}
For   given quantum numbers $n,l$  the string correction increases for larger
$l$, while for a fixed $l$ it decreases for higher radial excitations. For the
$1P$, $1D$ light mesons their values are typically equal to $\sim -40$~MeV,
$-70$~MeV, respectively (they were calculated using analytical expressions for
$\Delta_{\rm str}$ from \cite{ref.1},\cite{ref.2}, \cite{ref.24}).

The  nonperturbative SE correction to the quark (antiquark) mass
is of great importance to provide linear behavior of the Regge
trajectories \cite{ref.2}. As shown in Ref.~\cite{ref.10}, this
correction is flavor-dependent, depends on the averaged energy of
a quark, being very small for a heavy quark and large for a light
(strange) quark:
\begin{equation}
 \Delta_{\rm SE}=-\frac{3\sigma}{2\pi} \left(\frac{\eta_1}{\omega_1(nl)} -
 \frac{\eta_2}{\omega_2(nl)} \right).
\label{eq.06}
\end{equation}
The factor $\eta_f~(f=1,2)$ depends on the quark flavor  and the vacuum
correlation length:  $\eta_n=1.0$ for a light quark, $\eta_s=0.70$ for the $s$
quark. Notice that the number $3/2$ enters the SE term (\ref{eq.06}), instead
of the number 2 derived before in \cite{ref.10}; this change comes from  more
exact definition of the vacuum correlation length \cite{ref.25}.

From Eq.~(\ref{eq.06}) one can see that the averaged quark
energies $\omega_i$ play a special role: they determine both the
string and the SE contributions, and also enter all spin-dependent
potentials. In some potential models a negative overall constant
is often introduced in the mass term (which  plays the role of a
self-energy correction), however, such a constant violates the
linear behavior of the Regge trajectories.

The ``linear+GE" potential  $V_{\rm B}(r) $,  was already tested in a large
number of previous studies of heavy-light mesons \cite{ref.14} and
heavy-quarkonia \cite{ref.15}:
\begin{equation}
 V_{\rm B}(r)=\sigma r - \frac{4\alpha_{\rm B}(r)}{3 r},
\label{eq.08}
\end{equation}
where the vector coupling $\alpha_{\rm B}(r)$ is taken as in background
perturbation theory \cite{ref.26} with the freezing value $\alpha_{\rm
crit}=0.495 ~(n_f =3)$.

The vector coupling in coordinate space is defined through the
vector coupling $\alpha_B(q^2)$ in the momentum space:
\begin{equation}
\alpha_B(r) =\frac{2}{\pi}\int\limits_0^\infty
dq\frac{\sin(qr)}{q}\,\alpha_B(q), \label{eq.09}
\end{equation}
which is taken  in two-loop approximation,
\begin{equation} \alpha_B(q) =\frac{4\pi}{\beta_0t_B}\left(1-\frac{\beta_1}{\beta_0^2}
  \frac{\ln t_B}{t_B}\right).
\label{eq.10}
\end{equation}
Here the logarithm,
\begin{equation}
 t_B=\frac{q^2+M_B^2}{\Lambda_B^2},
\label{eq.11}
\end{equation}
contains the QCD constant  $\Lambda_B(n_f)$, which is defined via
the QCD constant $\Lambda_{\overline{MS}}(n_f)$ in the
$\overline{MS}$ scheme. The relation between them has been
established in \cite{ref.27}:
\begin{equation}
   \Lambda_B(n_f)=\Lambda_{\overline{MS}}\exp\left(-\frac{a_1}{2\beta_0}\right),
   \label{eq.12}
   \end{equation}
with $\beta_0=11 -\frac{2}{3}n_f$ and
$a_1=\frac{31}{3}-\frac{10}{9}n_f$. From the relation
(\ref{eq.12}) one can see that for a given $n_f$ the constant
$\Lambda_B(n_f)$ is significantly larger than
$\Lambda_{\overline{MS}}$:
\be
\nonumber \Lambda_B^{(5)}=1.3656\Lambda_{\overline{MS}}^{(5)}\quad (n_f=5);\\
\Lambda_B^{(3)}=1.4753\Lambda_{\overline{MS}}^{(3)}\quad (n_f=3). \label{eq.13}
\ee

At present only the QCD constant $\Lambda_{\overline{MS}}^{(5)}$ (for $n_f=5$)
is well known from experimental value of $\alpha_s(M_Z)=0.1182\pm 0.0012$; then
in two-loop approximation it gives
$\Lambda^{(5)}_{\overline{MS}}(\textrm{two-loop})=232(12)$~MeV. For $n_f=3$ the
QCD constant $\Lambda_{\overline{MS}}$ is not known  with a good accuracy and
to define it we fix here the freezing value $\alpha_{crit}$: $\alpha_{\rm
crit}(n_f=3) \simeq 0.495$. In (\ref{eq.11}) the background mass $M_B$ also
enters; its value is proportional to $\sqrt{\sigma}$ and for $\sigma=0.18$
GeV$^2$ the number $M_B=1.0\pm 0.05$~GeV was extracted from a detailed
comparison of the static force in the field correlator method used and in the
lattice QCD \cite{ref.28}  (here we take $M_B=1.0$~GeV).

The important feature of the critical couplings is that in the
momentum and coordinate space they coincide,
$\alpha_B(crit)=\alpha_B(q^2=0)= \alpha_B(r\rightarrow \infty)$:
\begin{equation}
   \alpha_B(crit)=\alpha_B (r\to \infty) =\alpha_B(q=0) =\frac{4\pi}{\beta_0
   t_0} \left( 1-\frac{\beta_1}{\beta_0^2}\frac{\ln
   t_0}{t_0}\right), \label{eq.14}
\end{equation}
with  $t_0= t_B(q^2=0)=ln\left(\frac{M_B^2}{\Lambda_B^2}\right)$. Thus in  our
calculations $\Lambda_B (n_f =3) =360$ MeV, $M_B =1.0$ GeV, $\alpha_{crit} =
0.4945,$ and $\sigma =0.18$ GeV$^2$.

In \cite{ref.2}, \cite{ref.16} it was shown that the GE
interaction remains important for the ground states of light, $K$,
and $\phi$ mesons, e.g. if GE interaction is neglected, then \be
\omega_1^{(0)} (1S) = \omega_2^{(0)}(1S) =335 ~{\rm MeV~for ~light
~mesons}\label{eq.15}\ee
$$\left. \begin{array}{l} \omega_1^{(0)}(1S)=347~{\rm Mev}\\\omega_2^{(0)} (1S) =411
\end{array}\right\} {\rm ~for ~K~ mesons},$$
while their values $\omega_1(nl),\omega_2(nl)$ increase if GE
interaction is taken into account (see Tables I,II). Thus for a
light meson  $\omega(1S)$ appears to be  $\sim 18\%$ larger and
such the growth of $\omega_i(nl)$ is important for more precise
definition of the magnetic moments of the $\rho$ and $K^*$ mesons.
Notice, that for higher excitations the GE interaction provides an
increase of $\omega_i(nl)$ by only $\sim 5\%$.



\end{document}